# A dissipative particle dynamics model of biofilm growth

Zhijie Xu[1,a], Paul Meakin[2,3,4], Alexandre Tartakovsky[1] and Timothy. D. Scheibe[5]

1.  Computational Mathematics Group, Fundamental and Computational Sciences Directorate, Pacific Northwest National Laboratory, Richland, WA 99352, USA

2. Center for Advanced Modeling and Simulation, Idaho National Laboratory, Idaho Falls, Idaho 83415, USA

3. Physics of Geological Processes, University of Oslo, Oslo 0316, Norway

4. Multiphase Flow Assurance Innovation Center, Institute for Energy Technology, Kjeller 2027, Norway

5. Hydrology Technical Group, Energy and Environment Directorate, Pacific Northwest National Laboratory, Richland, WA 99352, USA

a) Electronic mail: zhijie.xu@pnl.gov Tel: 509-372-4885

**Abstract**

A dissipative particle dynamics (DPD) model for the quantitative simulation of biofilm growth controlled by substrate (nutrient) consumption, advective and diffusive substrate transport, and hydrodynamic interactions with fluid flow (including fragmentation and reattachment) is described. The model was used to simulate biomass growth, decay, and spreading. It predicts how the biofilm morphology depends on flow conditions, biofilm growth kinetics, the rheomechanical properties of the biofilm and adhesion to solid surfaces. The morphology of the model biofilm depends strongly on its rigidity and the magnitude of the body force that drives the fluid over the biofilm.

**I. INTRODUCTION**

Dissipative particle dynamics (DPD) is a stochastic Lagrangian approach introduced by Hoogerbrugge and Koelman in 1992 [1]. DPD models are based on the idea that particles can be used to represent clusters of atoms or molecules instead of single atoms or molecules to provide a simple but robust way of coarse graining the molecular dynamics of dense fluid and soft condensed matter systems. DPD particle-particle interactions include conservative (non-dissipative) forces that arise from the conservative molecular interactions, and dissipative and fluctuating interactions, which are related by the fluctuation-dissipation theorem [2, 3]. The dissipative and fluctuating interactions originate from the internal degrees of freedom associated with individual DPD particles and together they function as a thermostat for the model. The grouping of atoms or molecules into a single DPD particle (coarse graining) leads to averaged effective conservative interaction potentials (purely repulsive potentials in the standard DPD model) between the DPD particles. Because the coarse grained DPD forces are much softer than the intermolecular and inter-atomic forces used in realistic molecular dynamics (MD) simulations, much longer time steps can be taken in DPD simulations, and the computational cost is substantially reduced. The coarse graining also reduces the computational burden, but this is less important than the time step increase. This makes DPD a very effective mesoscale particle simulation technique on length and time scales larger than those accessible to fully atomistic molecular dynamics (MD) simulations. DPD has been extensively used to investigate the effects of the size, shape and rigidity of large molecules, and their intermolecular interactions on the behavior of soft condensed matter [4-7]. In contrast to smoothed particle hydrodynamics (SPH), another popular particle model used in multiphase flow and reactive transport [8, 9], DPD provides a consistent way to simulate the effects of thermal fluctuations on biological systems such as lipid bilayers [10] which can be very important at small scales. Recent improvements in the implementation of no-slip boundary conditions for particle models make DPD a more accurate and efficient way of simulating fluid flow in geometrically complex confined systems [11, 12]. Hence, DPD is an attractive approach for the simulation of biological systems on supra-atomic length and time scales.

DPD has been used quite extensively to simulate the dynamic behavior of cell membranes, and lipid bilayers, which play very important roles in living cells [13-16]. Biofilm can be defined as a microbial community comprised of either a single or multiple species embedded in extracellular biopolymer that adheres to a solid substratum. Biofilm growth is important in medical applications including the development of tumors [17], the microbial contamination of prosthetic devices [18] and stents [19] and dental plaque [20]. Modeling the effects of nutrient and metabolic waste transport and fluid flow on biofilm development is challenging because of the complexity of the underlying coupled physical, chemical and biological processes that govern the evolution of biofilm structures over a wide range of scales [21]. Biofilm structure development can be conceptualized as a competition between "positive" processes that lead to biofilm volume expansion and "negative" processes [21] that lead to biofilm volume reduction. Typical "positive" processes include cell attachment, cell growth and division, in response to the transport of dissolved substrate via both advection and diffusion from or to the biofilm, and the secretion of extracellular polymeric substance (EPS). While typical "negative" processes include cell death, erosion (the removal of single cells or small cell clusters from the biofilm) and fracturing or sloughing (the removal of a large number of living and/or dead microoranisms in a single event) due to liquid-biofilm hydrodynamic interactions.

A variety of conceptual models, numerical models and numerical approaches have been used to simulate biofilm growth and behavior. A continuum model, similar to a phase-field model, which treats both the liquid and the biofilm phase as continuous media, but with different diffusion coefficients, $D_s$, was developed by Eberl, Parker and van Loosdrecht [22]. However, special attention must be given to the diffusion coefficient of the biofilm phase to account for the fact that biofilm spreading is significant only when the biomass density, $C_b$, is close to the maximum biomass density, $C_{bm}$. Grid-based cellular automata models have also been developed to simulate biofilm growth [21, 23, 24]. Knutson et al. used a lattice Boltzmann method to solve the Stokes equation for low Reynolds number fluid flow, a finite differences algorithm to solve the advection diffusion equation and a cellular automaton model for biofilm growth. This model takes into account the effect of fluid shear stresses on biomass growth [25]. Tartakovsky et al. used a smoothed particle hydrodynamics (SPH) method to solve governing transport equations while fluid-biofilm interactions were modeled via pair-wise short-range repulsive and medium-range attractive forces [26]. The hydrodynamic interactions between flow and biofilm structure can also be included in biofilm models by using a finite-element method to solve the stress and strain in the biofilm structure at each time step [24].

In contrast to grid-based computational methods, DPD, as a particle method, has the advantages of rigorous mass and momentum conservation and more importantly, explicit interface tracking is not needed for complex geometries and topological changes. In this paper we describe a DPD model for the growth and deformation of biofilm in a flowing fluid.

The hydrodynamic interactions between the biofilm and liquid flow are expected to be important in this application and they are included naturally in the DPD model. The DPD model of biofilm development provides important insight that contributes to a better understanding of the mechanisms of biofilm formation, growth and cell death.

## II. DISSIPATIVE PARTICLE DYNAMICS MODEL OF BIOFILM

### A. Mathematical model of biofilm

A mathematical model of biofilm development and behavior is needed to describe the associated underlying physical, chemical and biological processes in quantitative terms. The entire domain of interest is first divided into three phases, namely the liquid phase, the biofilm phase and the biofilm support (substratum) phase. Assuming that trace nutrients are present with adequate concentrations, the presence of two types of nutritional substrate, electron donors and electron acceptors, are generally required for biofilm growth. If there is an unlimited supply of one substrate (either electron donors or electron acceptors) then the biofilm growth is limited by the concentration of only one nutritional substrate, $S$ [21, 27]. The concentration of the nutritional substrate, $S$, and the biofilm growth, and are governed by the advection-diffusion-reaction equation:

$$\partial C_s / \partial t + \mathbf{V} \bullet \nabla C_s = D_s \nabla^2 C_s - r_s, \qquad (1)$$

where $V$ is the liquid velocity, $C_s(\mathbf{x}, t)$ is the concentration of substrate, $S$, at position $\boldsymbol{x}$ and time $t$, and $D_s$ is the diffusion coefficient of substrate $S$ in both liquid and biofilm phases (we assume equal diffusion coefficients in both phases). The substrate consumption rate in the biofilm phase, $r_s$, in units of $kg_s / (m^3 s)$, is the substrate consumption due to biofilm growth, where the subscript $s$ indicates a substrate variable or parameter. The subscript $b$ indicates a biomass variable or parameter and the subscript *bs* is used for parameters associated with biomass-substrate coupling). Following Picioreanu, van Loosdrecht, and Heijnen [27], we assume that for a single substrate, the consumption rate is given by the Monod function

$$r_s = \frac{\mu_m}{Y_{bs}} \frac{C_b C_s}{\left(K_s + C_s\right)},$$  (2)

where $\mu_m$ is the maximum biomass growth rate (s$^{-1}$), $K_s$ is the substrate saturation constant $\left(kg_s/m^3\right)$, and $Y_{bs}$ is the biofilm yield coefficient ($kg_b/kg_s$ the mass of biofilm that can be generated from a unit mass of substrate). $C_b(\mathbf{x}, t)$ is the biomass density in units of $kg_b/m^3$. The liquid velocity field, **V**, is given by the incompressible Navier-Stokes equations,

$$\nabla \bullet \mathbf{V} = 0,$$  (3)

$$\partial \mathbf{V} / \partial t + \mathbf{V} \bullet \nabla \mathbf{V} = -\nabla P/\rho + \nu \nabla^2 \mathbf{V}.$$  (4)

Equations (3) and (4) describe mass and momentum conservation in the liquid phase, where $P$ is the pressure field, $\rho$ is the liquid density and $\nu$ is the kinematic viscosity. The kinetic model describing the biofilm growth and/or decay in the biofilm phase is written as,

$$dC_b/dt = Y_{bs}\left(r_s - m_s C_b\right),$$  (5)

where $m_s$ is a maintenance coefficient with a unit of $kg_s/\left(kg_b \cdot s\right)$ representing the biomass decay effect. Biomass spreading is important in the biofilm kinetics model. The biomass density has a maximum value, $C_{bm}$, and whenever the biomass density, $C_b$, grows larger than the threshold value, $C_{bm}$, the extra biomass is redistributed giving rise to biofilm volume expansion. The original set of equations (1)-(5) can be rewritten in a dimensionless form by introducing the characteristic length, $l_c$, and velocity, $v_c$, (time $t^{'} = l_c/v_c$) as the units of length and velocity and substituting Eq. (2) into Eq. (5) leading to the equations

$$\partial c_s / \partial t' + \mathbf{v}' \bullet \nabla c_s = \nabla^2 c_s / P_e - k_1 k_4 \, c_b c_s / (k_2 + c_s), \tag{6}$$

$$\nabla \bullet \mathbf{v}' = 0, \tag{7}$$

$$\partial \mathbf{v}' / \partial t' + \mathbf{v}' \bullet \nabla \mathbf{v}' = -\nabla p' + \nabla^2 \mathbf{v}' / R_e, \tag{8}$$

$$dc_b / dt' = k_1 c_b \left[ c_s / (k_2 + c_s) - k_3 \right], \tag{9}$$

where the dimensionless numbers $k_1 - k_4$ are defined as, $k_1 = \mu_m l_c / v_c$, $k_2 = K_s / C_{sm}$,

$k_3 = Y_{bs} m_s / \mu_m$, and $k_4 = C_{bm} / (Y_{bs} C_{sm})$. $C_{sm}$ is the maximum substrate concentration in the system, and it is used for the purpose of normalization. The dimensionless substrate concentration and biomass density are normalized by $c_s = C_s / C_{sm}$ and $c_b = C_b / C_{bm}$, and they both lie in the range 0 to 1. The Reynolds number is defined as, $R_e = v_c l_c / \nu$ and the Péclet number is defined as $P_e = v_c l_c / D_s$. The dimensionless velocity and pressure fields are $\mathbf{v}' = \mathbf{V} / v_c$ and $p' = P / \rho v_c^2$. The dimensionless number $k_1$ is the ratio between the biofilm growth velocity and the characteristic fluid flow velocity. The dimensionless number $k_3$ is the ratio between the biofilm decay rate and the biofilm maximum growth rate, and $k_4$ is the ratio between the maximum biomass density and the biomass yield from the liquid with the maximum substrate concentration of $C_{sm}$. Further reduction of this set of dimensionless equations ((6)-(9)) by introducing a new time scale $t^* = t' k_1$ and velocity scale $\mathbf{v} = \mathbf{v}' / k_1$, gives

$$\partial c_s / \partial t^* + \mathbf{v} \bullet \nabla c_s = \nabla^2 c_s / P_{eb} - k_4 \, c_b c_s / (k_2 + c_s), \tag{10}$$

$$\nabla \bullet \mathbf{v} = 0, \tag{11}$$

$$\partial \mathbf{v} / \partial t^* + \mathbf{v} \bullet \nabla \mathbf{v} = -\nabla p + \nabla^2 \mathbf{v} / R_{eb}, \tag{12}$$

$$dc_b / dt^* = c_b \left\{ c_s / \left( k_2 + c_s \right) - k_3 \right\} , \qquad\qquad (13)$$

where the new set of equations contains only three dimensionless numbers, $k_2$, $k_3$, $k_4$, and two dimensionless constants, $R_{eb} = \mu_m l_c^2 / \nu$ and $P_{eb} = \mu_m l_c^2 / D_s$. The corresponding dimensionless time, velocity, and pressure fields are defined as $t^* = \mu_m t$, $\mathbf{v} = \mathbf{V} / \left( \mu_m l_c \right)$ and $p = P / \rho \left( \mu_m l_c \right)^2$. Some relevant time scales in the DPD model can be identified from Eqs. (10)-(13). For example, the time scale for biofilm decay is $\tau_{bd} = 1 / k_3$ and the time scale for biofilm growth is $\tau_{bg} = \left( k_2 + c_s \right) / c_s$, with a range of $\left( k_2 + 1 \right) \leq \tau_{bg} < \infty$, depending on the substrate concentration, $c_s$. A lower substrate concentration, $c_s$, leads to slower biofilm growth and a longer growth time scale, $\tau_{bg}$. The critical substrate concentration required for biomass to keep growing, obtained from Eq. (13) by equating the time scales $\tau_{bg}$ and $\tau_{bd}$ ( $\tau_{bg} = \tau_{bd}$ ) is $c_s^c = k_2 k_3 / \left( 1 - k_3 \right)$. The time scale associated with the substrate diffusion is $\tau_d = P_{eb}$ and the time scale associated with consumption (or reaction) is $\tau_c = \left( k_2 + c_s \right) / \left( k_4 c_b \right)$, with a range of $k_2 / k_4 \leq \tau_c$. The time scales for substrate advection and fluid momentum diffusion can be defined as $\tau_a = 1 / \mathbf{v}$ and $\tau_m = R_{eb}$. Biofilm growth is an intrinsically multiscale phenomenon characterized by multiple time and length scales. The balance between substrate transport and consumption in Eq. (10) determines the substrate concentration level $c_s$, which in turn controls the biofilm growth rate through Eq. (13). This makes the problem of biofilm growth and decay similar to mineral precipitation and/or dissolution problems [28, 29]. By making the time scale of substrate diffusion comparable to the shortest substrate consumption time scale ( $\tau_d = k_2 / k_4$ ), a dimensionless number (similar to the Damköhler number $D_a$ in a mineral precipitation and/or dissolution)

$$GR = k_4 P_{eb} / k_2 = \frac{\mu_m l_c^2}{D_s} \cdot \frac{C_{bm}}{Y_{bs} K_s} \qquad (14)$$

can be introduced, and this dimensionless ratio has an important influence on biofilm growth. The value of $GR$ characterizes two growth regimes, with a low value of $GR$ corresponding to the reaction limited regime and a high value of $GR$ corresponding to the diffusion limited regime. Appropriate velocity, concentration and pressure boundary conditions associated with Eqns. (10)-(13) must be provided to complete the definition of the model.

## B. Dissipative particle dynamics biofilm model

Standard dissipative particle dynamics (DPD) uses an ensemble of particles to represent fluids [1]. DPD particles move due to the combination of conservative (non-dissipative), $\mathbf{f}^C$, dissipative, $\mathbf{f}^D$, fluctuating (random) $\mathbf{f}^R$, and external, $\mathbf{f}^{ext}$, forces. The equation of motion for the DPD particles is

$$m_i d\mathbf{v}_i / dt = \mathbf{f}_i^{int} + \mathbf{f}_i^{ext} = \mathbf{f}_i^C + \mathbf{f}_i^D + \mathbf{f}_i^R + \mathbf{f}_i^{ext} = \mathbf{f}_i^{ext} + \sum_{j \neq i} (\mathbf{f}_{ij}^C + \mathbf{f}_{ij}^D + \mathbf{f}_{ij}^R), \qquad (15)$$

where $\mathbf{v}_i$ is the velocity of particle $i$ and $m_i$ is its mass. In models for single phase fluid flow, the conservative forces between particles are usually given by a simple purely repulsive form such as $\mathbf{f}_{ij}^C = S(1 - r_{ij} / r_0)\hat{\mathbf{r}}_{ij}$ for $r_{ij} = |\mathbf{r}_{ij}| = |\mathbf{r}_i - \mathbf{r}_j| < r_0$ and $\mathbf{f}_{ij}^C = 0$ for $r_{ij} \geq r_0$, where $S$ is the strength of the particle-particle interaction, $r_0$ is the cutoff range of the particle-particle interactions, and $\hat{\mathbf{r}}_{ij}$ is the unit vector pointing from particle $j$ to particle $i$ ($\hat{\mathbf{r}}_{ij} = (\mathbf{x}_i - \mathbf{x}_j) / |\mathbf{x}_i - \mathbf{x}_j|$). The total conservative force acting on particle $i$ is the sum of the conservative forces between particle i and all other neighboring particles within the cutoff distance,

$$\mathbf{f}_i^C = \sum_{j \neq i} \mathbf{f}_{ij}^C = \sum_{j \neq i, r_{ij} < r_0} S_{ij}(1 - r_{ij}/r_0)\hat{\mathbf{r}}_{ij}, \tag{16}$$

where $S_{ij}$ is the strength of the interaction between particle $i$ and particle $j$. The dissipative particle-particle interactions are given by $\mathbf{f}_{ij}^D = \gamma W^D(r_{ij})(\mathbf{r}_{ij} \cdot \mathbf{v}_{ij})\hat{\mathbf{r}}_{ij}$, where $\gamma$ is a viscosity coefficient and $\mathbf{v}_{ij} = \mathbf{v}_j - \mathbf{v}_i$, for $r_{ij} < r_0$ and $\mathbf{f}_{ij}^D = 0$ for $r_{ij} > r_0$ so that

$$\mathbf{f}_i^D = \sum_{j \neq i} \mathbf{f}_{ij}^D = -\sum_{j \neq i, r_{ij} < r_0} \gamma W^D(r_{ij})(\mathbf{r}_{ij} \cdot \mathbf{v}_{ij})\hat{\mathbf{r}}_{ij}. \tag{17}$$

The random force is given by $\mathbf{f}_{ij}^R = \sigma W^R(r_{ij})\zeta\hat{\mathbf{r}}_{ij}$ for $r_{ij} < r_0$ and $\mathbf{f}_{ij}^R = 0$ for $r_{ij} > r_0$, where $\sigma$ is the fluctuation strength coefficient and $\zeta$ is a random variable selected from a Gaussian distribution with a zero mean and a unit variance so that

$$\mathbf{f}_i^R = \sum_{j \neq i} \mathbf{f}_{ij}^R = \sum_{j \neq i, r_{ij} < r_0} \sigma W^R \zeta\hat{\mathbf{r}}_{ij}. \tag{18}$$

The random and dissipative particle-particle interactions are related through the fluctuation-dissipation theorem [3], which requires that $\gamma = \sigma^2/2k_B T$. Here, $k_B$ is the Boltzmann constant, $T$ is the prescribed DPD simulation temperature, and $W^D(r) = (W^R(r))^2$, where $W^D(r)$ and $W^R(r)$ are $r$-dependent weight functions, both vanishing for $r \geq r_0$. In standard DPD models, simple weighting functions defined as $W^D(r) = \left[W^R(r)\right]^2 = \left(1 - r/r_0\right)^2$ (for $r < r_0$) are used. The combination of dissipative and fluctuating forces, related by the fluctuation-dissipation theorem [2] acts as a thermostat, which maintains the temperature of the system, measured through the average kinetic energy of the particles at a temperature of $T$, provided that the time step used in the simulation is small enough.

In the current DPD biofilm model, three types of DPD particles are used to represent the three distinct phases (liquid, biofilm, and substratum). The solution of the Navier-Stokes equations (Eqs. (11) and (12)) is approximated by the low Mach number flow of a slightly compressible fluid represented by the DPD particles. The velocities and positions of the DPD particles are found from Eq. (15) using a modified velocity Verlet algorithm to integrate the equation of motion [30]. The liquid-biofilm and liquid-substratum particle-particle interactions should be strong enough to prevent the penetration of the liquid particles into the biofilm and substratum regions and this simulates the no-slip boundary condition on the liquid-biofilm and liquid-substratum interfaces.

The masses of substrate and biomass carried by DPD particle $i$ are specified by the substrate concentration, $c_{s,i}$, and the biomass density, $c_{b,i}$; and the changes in the substrate concentrations in the liquid and biofilm DPD particles are given by the advection-diffusion-reaction equation (Eq. (10)). The DPD representation of Eq. (10)

$$\frac{dc_{s,i}}{dt} = \sum_{j \neq i} \lambda_{ij} W^R(r_{ij}) \left( c_{s,j} - c_{s,i} \right) + \sum_{j \neq i} \zeta^c \sqrt{2\alpha^{-1}\lambda_{ij} W^R(r_{ij}) c_{s,i} c_{s,j} / \Delta t} - \frac{k_s c_{b,i} c_{s,i}}{k_2 + c_{s,i}}, \qquad (19)$$

is similar to the DPD representation of the heat conduction equation [31-33], where $\lambda_{ij}$ is the inter-particle diffusion constant between particles $i$ and its neighbor $j$, which can be related to the continuum molecular diffusion coefficient. A simple mean field theory calculation yields an *a priori* estimate for the substrate diffusivity in the DPD model system, $D_s^{DPD} \propto \rho_{DPD} \lambda_{ij} r_c^2$, where $\rho_{DPD}$ is the particle density and $r_c$ is the cut off distance [33]. The first term on the RHS side of Eq. (19) is a dissipative term that represents the diffusive exchange of substrate between neighboring DPD particles due to the concentration differences. It also takes into account the random exchange of substrate between DPD particles due to thermally induced concentration fluctuations through the second term. $\Delta t$ is the time step used in the DPD simulations, $\alpha$ is a material constant representing the magnitude of the concentration fluctuations, $\zeta^c$ is a random variable of the same type as $\zeta$ in Eq. (18), but it is uncorrelated with $\zeta$. The reason why the factor of $\Delta t^{-1/2}$ is used in Eq. (19) is because the average of a random fluctuation acting over a time step of length $\Delta t$ is proportional to $\Delta t^{1/2}$, and this can be represented as a randomly selected constant with a magnitude proportional to $\Delta t^{-1/2}$ acting over a time of $\Delta t$ [30]. The DPD equation for evolution of the biomass density obtained from Eq. (13) is:

$$dc_{b,i}/dt = c_{b,i}\left\{c_{s,i}/\left(k_2 + c_{s,i}\right) - k_3\right\}. \tag{20}$$

For the DPD particles that represent biofilm $c_{b,i} > 0$, and $c_{b,i} = 0$ for all liquid and substratum DPD particles. Biomass spreading is important for biofilm volume expansion. If any DPD particle has a biomass density, $c_{b,i}$, that exceeds the maximum biomass density $c_{bm}$ (a normalized biomass density of unity), the excess biomass is instantaneously transferred to the nearest fluid DPD particle within the cutoff distance and this fluid DPD particle is spontaneously changed to a biofilm type DPD particle. The excess biomass is assumed to be lost if there are no fluid DPD particles within the cutoff range. In practice, biomass grows primarily near interface between the biomass and the fluid, and it decays in the interior of the biomass domain. If a DPD biofilm particle does not have any fluid particles within the cutoff range, it is very probably well inside the biomass domain where the biomass is decaying. Therefore, it is very unlikely that excess biomass will be discarded in a simulation. This biomass spreading algorithm is similar to the discrete rules used to redistribute biomass in some cellular automata models [23], where a search for "free-space" among the nearest-neighbor elements is also carried out.

## III. RESULTS AND DISCUSSION

### A. Non-deformable biofilm growth

2D DPD simulations of biofilm growth and fluid flow with ~6000 DPD particles in a narrow channel with a width of 20 (half width of $a = 10$) and length of 50 were performed. Periodic boundary conditions were applied along the flow direction. The density of the fluid was set to $\rho = 4$, and the prescribed temperature for the DPD simulations was set to $k_B T = 1.0$. The dissipative and random coefficients were chosen to be $\gamma = 4.5$ and $\sigma = 3.0$ to satisfy the fluctuation-dissipation theorem and the conservative force parameter between fluid-fluid particles was set to $S_{ff} = 18.75$. The DPD parameters were chosen to match the compressibility of water [30]. The fluid flow calculations have been validated in the previous work [11]. Fluid-wall (or substratum), fluid-biofilm interaction was $S_{fw} = S_{fb} = 18.75$ - the same as the fluid-fluid interaction. Here the subscripts indicate the various types of DPD particle-particle interactions (*ff* represents fluid-fluid interaction, *fw* represents fluid-substratum interaction, and *fb* represents fluid-biofilm interaction). The entire domain was divided into $60 \times 30$ bins and data was collected and averaged for each bin to obtain the average flow field. Stable flow and concentration fields were first established and then a certain number of biofilm particles (8 in the current simulations) were inoculated on the lower substratum surface to represent the initial adhesion of microbes. The biomass starts to grow due to absorption of substrate from the surrounding liquid. Periodic boundary conditions were applied along the flow direction: a fluid particle exiting the right boundary is reinserted into the computational domain through the left boundary with a fixed substrate concentration of 1.0 prescribed at the boundary.

The Reynolds number $R_{eb}$ in Eq. (12) is related to the kinematic viscosity $\nu^{DPD}$ of the DPD fluid under investigation $R_{eb} = 1/\nu^{DPD}$, with $\nu^{DPD} = 0.23$ determined from previous work.[11] The Péclet number $P_{eb}$ in Eq. (10) should be related to the dimensionless diffusion coefficient $D_s^{DPD}$ of the DPD fluid by $P_{eb} = 1/D_s^{DPD}$. The DPD diffusion coefficient, $D_s^{DPD}$, is approximately related to the inter-particle diffusion constant $\lambda_{ff}$ by $D_s^{DPD} \approx 3\rho\lambda_{ff}/40$ for 2D simulations [33]. In principle, there are two contributions to $D_s^{DPD}$: one is from the exchange of solute between adjacent DPD particles, and the other from the random motion of the DPD particles. The effective diffusion coefficient is the sum of the two contributions. However, the second contribution is negligible for large $D_s^{DPD}$. An external body force of magnitude $g$, acting along the channel axis, was imposed on each fluid particle to initiate and sustain the flow. By varying the magnitudes of $\lambda_{ff}$ and $g$, DPD fluids with various diffusion coefficients and flow velocities can be generated. In the current simulations, a value of $\lambda_{ff} = 3.3$ (or equivalently $D_s^{DPD} \approx 1$ and $P_{eb} \approx 1$) and $g = 0.02$ were used as a reference. The dimensionless numbers $k_2 = 0.0875$ and $k_3 = 0.09$ were used in all simulations, and various values for the constant $k_4$ were selected to simulate different biofilm growth regimes.

For the sake of simplicity, the biofilm structures were first assumed to be non-deformable, and the positions of the biofilm particles were fixed (once a liquid DPD particle is transformed into a biofilm DPD particle due to local biomass spreading, it is not allowed to move). Figures 2-4 show the time development of non-deformable biofilm structures simulated with $k_4 = 2.0$, 5.0 and 10.0. The gravity driven fluid flow is from left to right, with red (dark gray) representing the substratum (channel wall), blue (light gray) representing the flowing liquid and green (white) representing the biofilm phase. The flow velocity vector field is indicated by black arrows, and the concentration field is represented by white contours from 0.0 to 1.0 with increments of 0.1. A smaller $k_4$ (or smaller $GR$) leads to faster growth in the biofilm phase and a compact biofilm structure (characteristic of the reaction-limited regime) was developed (Fig. 2). The separate biofilm clusters grow and merge into a large continuous biofilm. At larger $k_4$ or $GR,$ biofilm growth becomes slower, and a more sprawling structure, characteristic of diffusion-limited growth, is formed (Figs. 3 and 4). The biofilm growth becomes biased towards the bulk liquid because the biofilm consumes substrate and the substrate concentration is highest near the most exposed upstream parts of the biofilm.

Substrate necessary for biofilm growth must diffuse through a thin layer in which the substrate concentration gradient is high to reach the biofilm cells. This concentration boundary layer is closely correlated with the hydrodynamic boundary layer generated by the no-slip boundary conditions and hydrodynamic screening of the recessed parts of the biofilm-fluid interface by those parts that protrude into the flowing fluid. The concentration contours in Figs. 2-4 reveal that the thickness of the concentration boundary layer (for example the separation distance between concentration contours from 0.0 to a small concentration, $\delta$ on the order of 0.1) increases with $k_4$, and a steep concentration gradient was found for fast biofilm growth at smaller $k_4$. The transition from compact growth for small values of $k_4$ to more extended growth for larger values of $k_4$ is similar to the transition from compact reaction-limited precipitation to dendritic diffusion-limited precipitation as the rate of the interface kinetics increases and the growth rate becomes limited by the rate of transport of solute to the interface. The thickness of the concentration boundary layer determines the resistance to mass transfer in the layer and a thinner boundary layer promotes the diffusive substrate transport from the liquid to biofilm phase and leads to faster growth and a more compact and smooth biofilm structure.

Figure 5 shows the biomass density distribution corresponding to the last snapshot (d) in Figs. 2-4. This figure shows only the "live" biomass (it does not take into account the organic material that remains after live biomass decays). An interesting finding is that the thickness of the biofilm layer that has nonzero live biomass density increases with $k_4$. For a smaller $k_4$, substrate can penetrate further into the biofilm and biomass can grow in a thicker layer, while for a large $k_4$, substrate can penetrate only a small distance into the biofilm, and the live biomass is concentrated into a thin layer near the surface. In this situation the internal mass diffusion inside the biofilm phase has no significant effect on the substrate conversion. Under these conditions, biofilm growth is similar to a surface reaction process in which the total substrate absorption is proportional to the total surface area of the biofilm phase [21]. A detailed analysis indicates that the thickness of the layer into which substrate can penetrate is related to the characteristic substrate diffusion length in the biofilm phase which is given by $h \approx \sqrt{2 D_s^{DPD} \tau_c} = \sqrt{2/GR}$, where $\tau_c$ is the time scale of substrate consumption.

The effect of flow velocity on biofilm development is illustrated in Fig. 6. The average flow velocity in the channel was changed by changing the gravitational body force, $m_i g$, that is applied to each liquid DPD particle to drive the liquid through the channel. Substrate transport is enhanced by increasing the flow velocity, and biofilm branches situated further from the flow inlet can obtain more substrate and grow much faster. This leads to a more compact biofilm structure. In general, the overall growth rate is higher for larger flow velocities The thickness of the concentration boundary layer also decreases with increasing flow velocity, which results in an increased substrate concentration gradient near to the biofilm surface and enhancement of the substrate mass transfer into the biofilm phase. However, a higher flow velocity leads to higher shear stresses on the biofilm structure and within the biofilm.

**B. Deformable biofilm growth**

In order to model deformable biofilm, an interaction potential between biofilm DPD particles is needed. Simple Lennard-Jones, biharmonic, and finite extensible nonlinear elastic (FENE) potentials have a long history in molecular dynamics studies of solids [34, 35]. In this work, a simple harmonic potential with an equilibrium distance, $r_e$, that is smaller than the cutoff distance, $r_0$, at which the force falls abruptly to zero and the "bond" between the two particles is effectively ruptured, was used to model the interactions between biofilm DPD particles, and this results in a soft solid-like biofilm structure. The harmonic potential energy and force between biofilm particles $i$ and $j$ are given by:

$$e_{ij} = -\frac{S_{bb}}{2r_e}(r_{ij} - r_e)^2,$$ (21)

$$\mathbf{f}_{ij} = -\mathbf{f}_{ji} = S_{bb}(1 - r_{ij}/r_e)\hat{\mathbf{r}}_{ij} \text{ for } r_{ij} < r_0,$$ (22)

$$\mathbf{f}_{ij} = 0 \text{ for } r_{ij} > r_0$$

The biofilm particle interaction strength, $S_{bb}$, and the equilibrium distance, $r_e$, control the mechanical properties of the biofilm. In the model, the interaction between a liquid DPD particle and a biofilm DPD particle is assumed to be equal to the liquid-liquid interaction ( $S_{fb} = S_{ff}$ , with purely repulsive interactions between liquid and biofilm particles). For a given geometry and gravitational acceleration, the rigidity of the biofilms is controlled by the magnitude of biofilm-biofilm DPD particle-particle interactions. The structure of the deformable biofilm responds to changing flow conditions. Two typical scenarios for deformable biofilm structures incorporating nutrient substrate transport, biofilm growth, and hydrodynamic interactions are shown in Fig. 7. The biofilm-biofilm interaction strength was first set to $S_{bb} = 2S_{ll}$ and the gravitational force was set to $g = 0.005$. The biofilm structure was strongly deformed and stretched in the fluid flow direction by the shear stress exerted by the flowing liquid at the liquid-biofilm interface, and biofilm bodies can detach very easily as seen in Fig. 7. This leads to a flat and smooth biofilm structure. For a deformable biofilm, the arms of the structures grown at large *GR* stretch and bend in the liquid, and eventually break at high flow velocity. At an even larger flow velocity, larger stresses at the biofilm-substratum interface can lead to detachment of the entire biofilm body from the substratum. This sloughing process can significantly change the biofilm morphology.

## V. CONCLUSIONS

A dissipative particle dynamics model for biofilm structure formation has been developed. The model incorporates fluid flow, substrate diffusion and advection, biofilm growth, and/or deformation. In DPD model, no interface tracking is needed to simulate the complex interface dynamics associated with biofilm growth, biofilm deformation and the advection of biofilm fragments. A two-dimensional version of the model was used to investigate biofilm growth in a narrow channel. The simulation results demonstrate the effects of flow velocity, growth parameter, and hydrodynamic interaction on the biofilm growth regime and morphology. It would be straightforward to extend the model to three spatial dimensions, but the simulations would require substantially more computational effort.

**ACKNOWLEDGMENTS**

This work was supported by the U.S. Department of Energy, Office of Science Scientific Discovery through Advanced Computing Program. The Idaho National Laboratory is operated for the U.S. Department of Energy by the Battelle Energy Alliance under Contract DE-AC07-05ID14517.

Figure 1. A schematic representation of the DPD biofilm model. Open circles represent liquid DPD particles flowing through a channel, filled circles indicate the immobile DPD particles used to represent the solid substratum and partially filled circles represent the biofilm DPD particle.

Figure 2. (Color online) DPD simulation of the growth of non-deformable biofilm for $k_4 = 2.0$ and $g = 0.02$. The snapshots were taken at time $t$ = 30, 60, 90, 120. White lines are isoconcentration contours with increment of 0.1 ranging from 0 (close to biofilm) to 1.0 (close to inlet). Black arrows are the flow velocity vectors and scaled by the magnitude.

Figure 3. (Color online) DPD simulation of the growth of non-deformable biofilm for $k_4 = 5.0$ and $g = 0.02$. The snapshots were taken at time $t$ = 60, 120, 180, 240. White lines are isoconcentration contours with increment of 0.1 ranging from 0 (close to biofilm) to 1.0 (close to inlet). Black arrows are the flow velocity vectors and scaled by the magnitude.

Figure 4. (Color online) DPD simulation of the growth of non-deformable biofilm for $k_4 = 10.0$ and $g = 0.02$. The snapshots were taken at time $t$ = 200, 300, 400, 500. White lines are isoconcentration contours with increment of 0.1 ranging from 0 (close to biofilm) to 1.0 (close to inlet). Black arrows are the flow velocity vectors and scaled by the magnitude.

Figure 5. (Color online) Plot of biomass density ($C_b$) spatial distribution corresponding to the last snapshots (d) in Figures 2, 3, and 4. Wall particles are not shown for clarity. Red color

(white) represents the highest biomass density and blue (dark gray) color represents the lowers biomass density.

Figure 6. (Color online) DPD simulation of the growth of non-deformable biofilm for $k_4 = 5.0$ at various flow velocities with $g = 0.005$, 0.02, and 0.05. White lines are isoconcentration contours with increment of 0.1 ranging from 0 (close to biofilm) to 1.0 (close to inlet). Black arrows are the flow velocity vectors and scaled by the magnitude.

Figure 7. (Color online) DPD simulation of the growth of deformable biofilm for $k_4 = 5.0$ with $g = 0.005$ shows the detachment of a part of biofilm body from bulk biofilm structure.

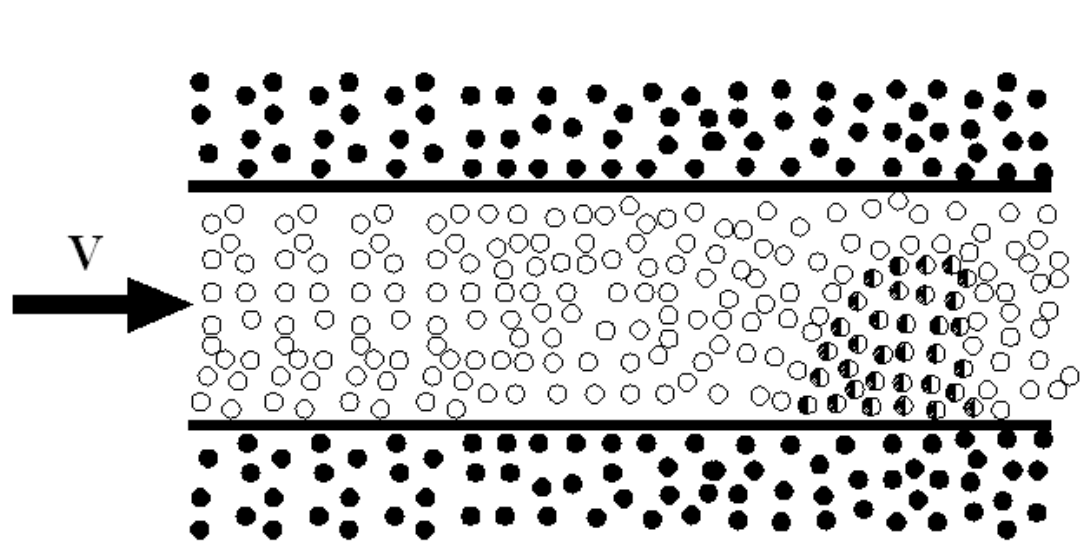

FIG. 1.

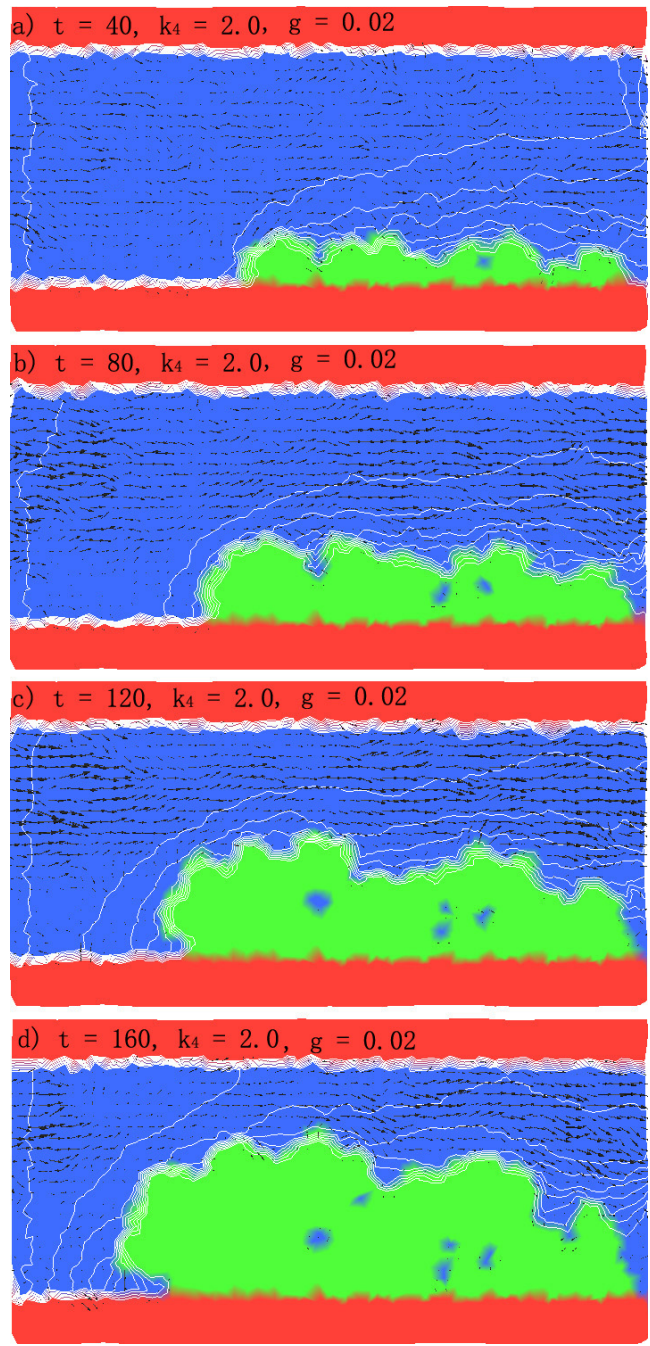

a) t = 40, k₄ = 2.0, g = 0.02

b) t = 80, k₄ = 2.0, g = 0.02

c) t = 120, k₄ = 2.0, g = 0.02

d) t = 160, k₄ = 2.0, g = 0.02

FIG. 2

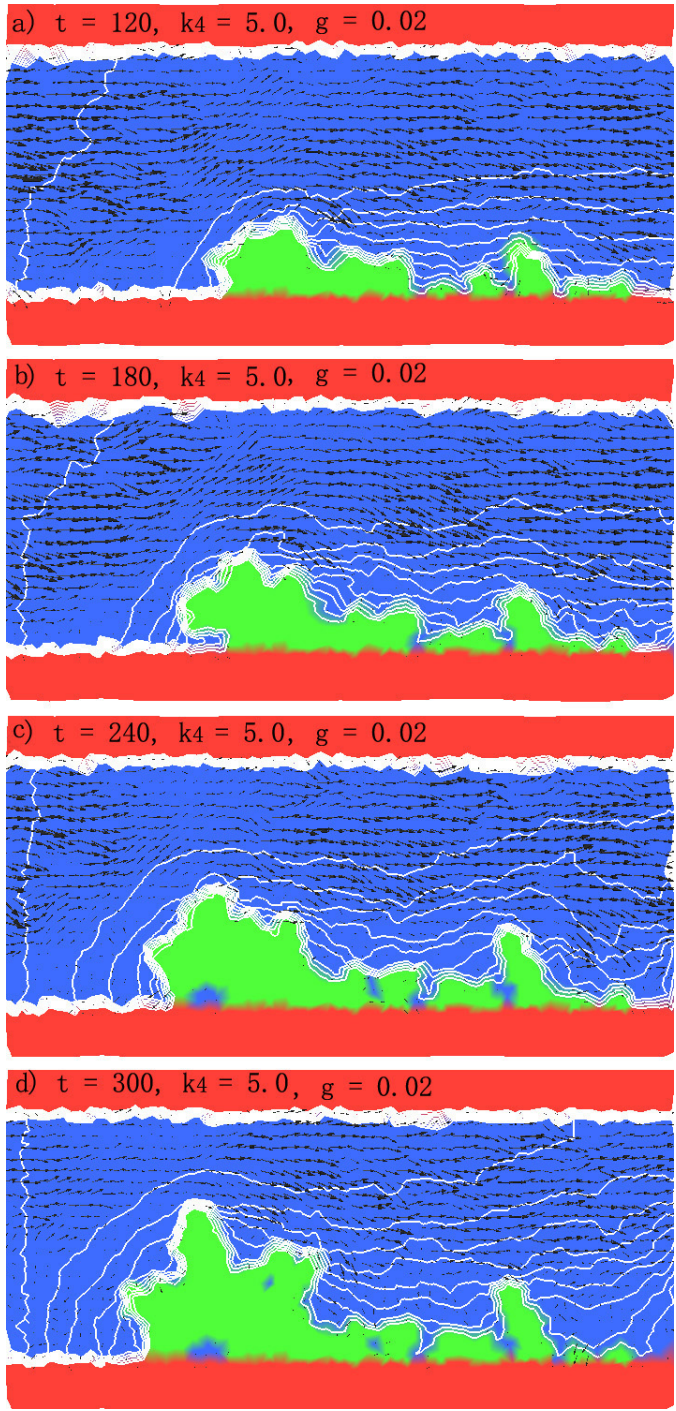

a)  t = 120,  k4 = 5.0,  g = 0.02

b)  t = 180,  k4 = 5.0,  g = 0.02

c)  t = 240,  k4 = 5.0,  g = 0.02

d)  t = 300,  k4 = 5.0,  g = 0.02

FIG. 3

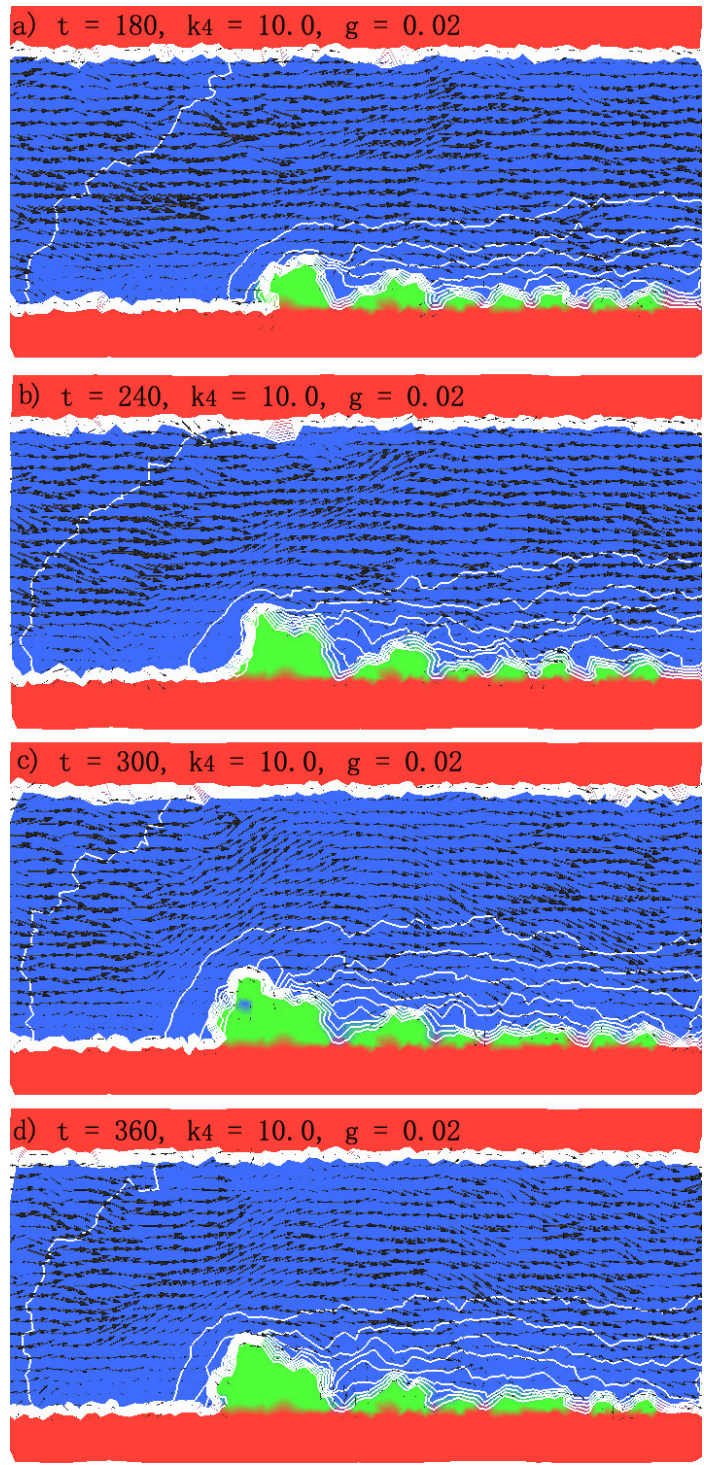

FIG. 4

a) Figure 2 (d)

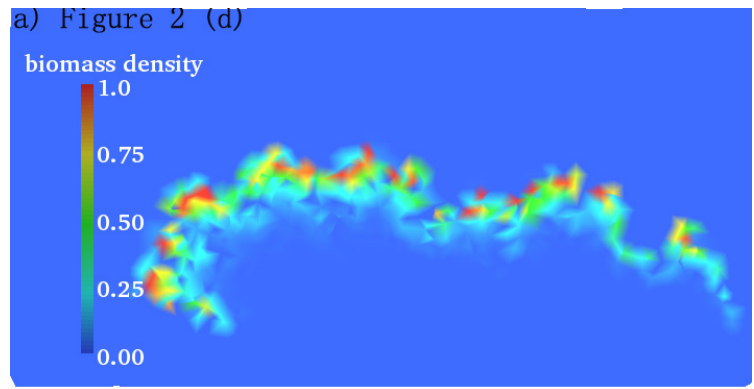

b) Figure 3 (d)

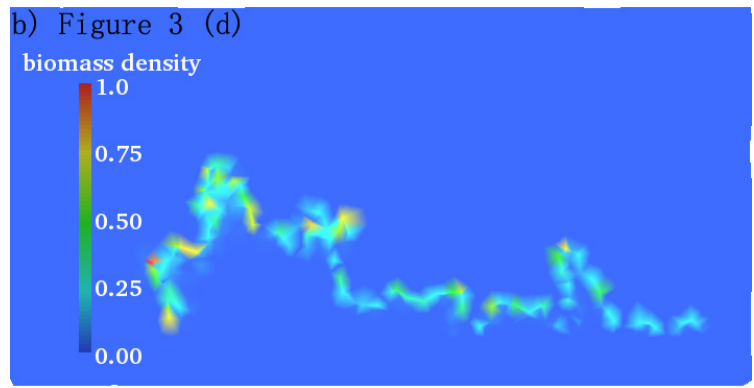

c) Figure 4 (d)

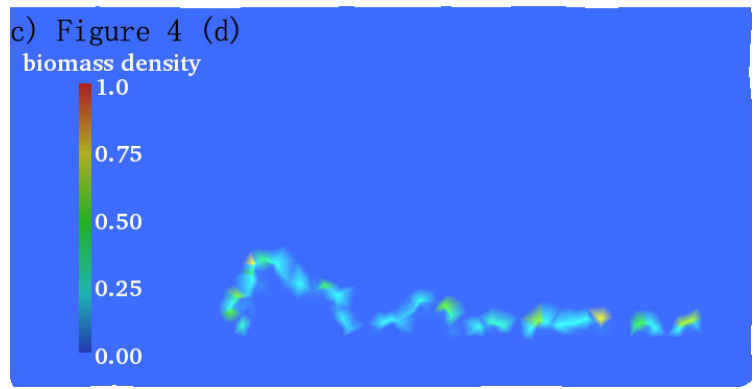

FIG. 5

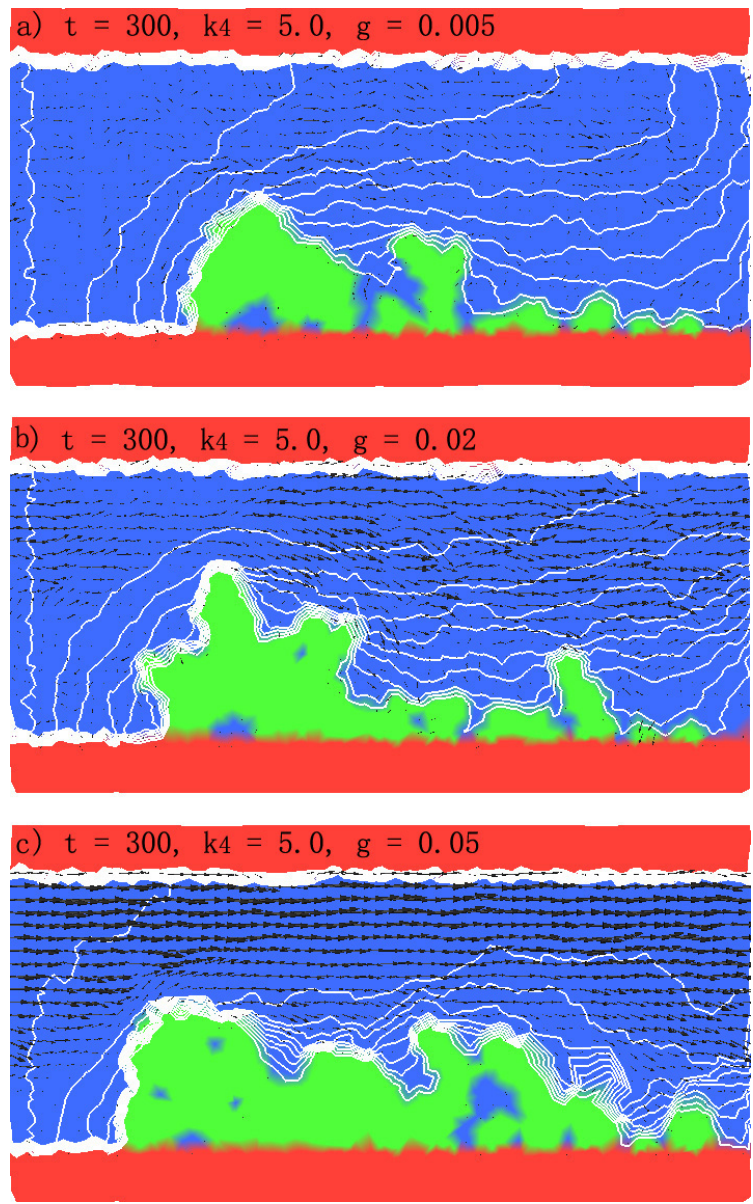

a) t = 300, k4 = 5.0, g = 0.005

b) t = 300, k4 = 5.0, g = 0.02

c) t = 300, k4 = 5.0, g = 0.05

FIG. 6

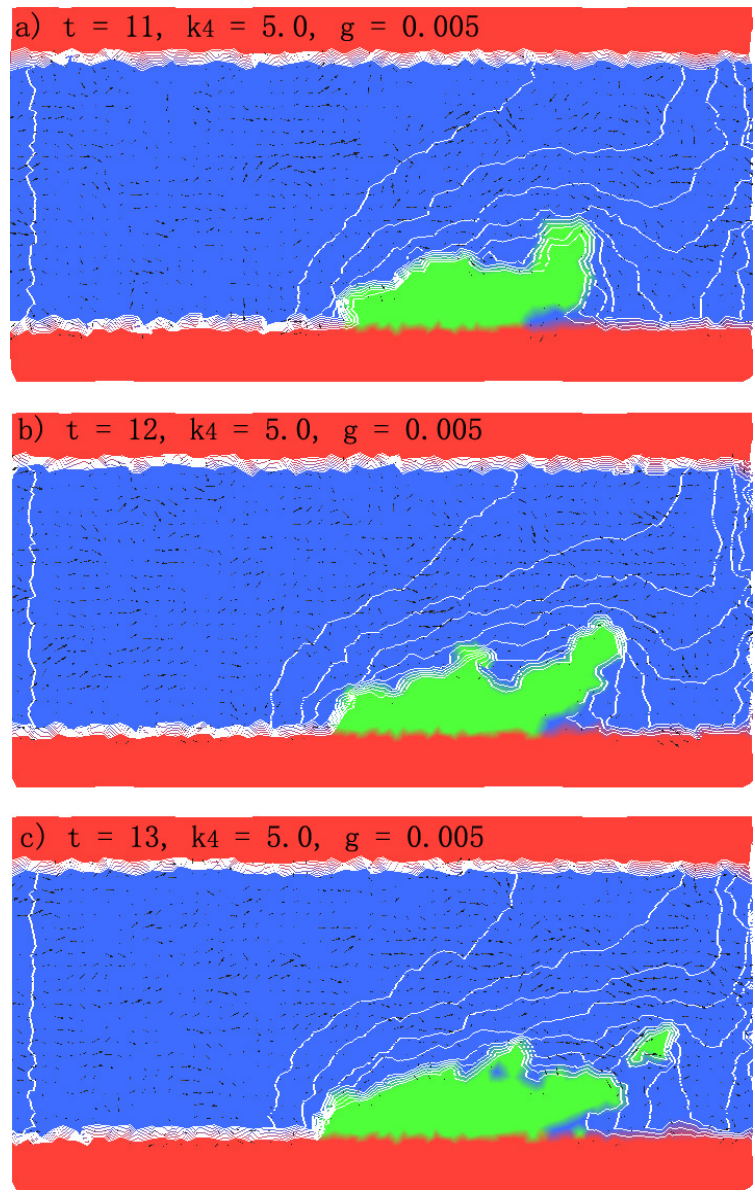

FIG. 7.